\documentclass[usenatbib,a4paper,fleqn]{mnras}

\usepackage[T1]{fontenc}
\usepackage{ae,aecompl}

\usepackage{newtxtext,newtxmath}

\usepackage{graphicx}	
\usepackage{amsmath}	
\usepackage{amssymb}	
\usepackage{multicol}   

\newcommand{\be}{\begin{equation}}
\newcommand{\ee}{\end{equation}}
\newcommand{\bea}{\begin{eqnarray}}
\newcommand{\eea}{\end{eqnarray}}

\newcommand{\revised}[1]{#1}
\newcommand{\changed}[1]{}

\newcommand{\m}{\,\hbox{m}}
\newcommand{\mm}{\,\hbox{mm}}
\newcommand{\km}{\,\hbox{km}}

\newcommand{\AU}{\,\hbox{AU}}
\newcommand{\g}{\,\hbox{g}}

\newcommand{\Myr}{\,\hbox{Myr}}
\newcommand{\Gyr}{\,\hbox{Gyr}}

\newcommand{\erg}{\,\hbox{erg}}

\newcommand{\kg}{\,\hbox{kg}}
\newcommand{\K}{\,\hbox{K}}
\newcommand{\total}{\,\hbox{d}}
\newcommand{\Mearth}{\,M_{\earth}}
\newcommand{\Msun}{\,M_{\sun}}
\newcommand{\Lsun}{\,L_{\sun}}

\newcommand{\AAp}      {Astron. Astrophys.}

\newcommand{\AJ}       {Astron. J.}

\newcommand{\ApJ}      {Astrophys. J.}
\newcommand{\ApJL}      {Astrophys. J. Lett.}
\newcommand{\ApJS}     {Astrophys. J. Suppl.}

\newcommand{\ARAA}     {Ann. Rev. Astron. Astrophys.}

\newcommand{\MNRAS}    {MNRAS}

\newcommand{\PASP}     {PASP}

\newcommand{\QJRAS}    {Quart. J. Roy. Astron. Soc.}

\title[Two-component debris discs]{Does warm debris dust stem from asteroid belts?}

\author[Geiler \& Krivov]{
        Fabian Geiler $^{1}$ \&
        Alexander V. Krivov $^{1}$
       \\
$^{1}$ Astrophysikalisches Institut und Universit\"atssternwarte, 
Friedrich-Schiller-Universit\"at Jena,
           Schillerg\"a{\ss}chen~2--3, 07745 Jena, Germany
          }

\date{Received {\em 16 December 2016}; accepted {\em 20 February 2017}}

\pubyear{2017}

\begin{document}
\pagerange{\pageref{firstpage}--\pageref{lastpage}}
\label{firstpage}
\maketitle
\begin{abstract}
Many debris discs reveal a two-component structure, with a cold outer and a warm inner component. While the former are likely massive analogues of the Kuiper belt, the origin of the latter is still a matter of debate. In this work we investigate whether the warm dust may be a signature of asteroid belt analogues. In the scenario tested here the current two-belt architecture stems from an originally extended protoplanetary disc, in which planets have \revised{opened a gap} separating it into the outer and inner discs which, after the gas dispersal, experience a steady-state collisional decay. This idea is explored with an analytic collisional evolution model for a sample of 225 debris discs from a Spitzer/IRS catalogue that are likely to possess a two-component structure. We find that the vast majority of systems (220 out of 225, or 98\%) are compatible with this scenario. For their progenitors, original protoplanetary discs, we find an average surface density slope of $-0.93\pm0.06$ and an average initial mass of $\left(3.3^{+0.4}_{-0.3}\right)\times 10^{-3}$ solar masses, both of which are in agreement with the values inferred from submillimetre surveys. However, dust production by short-period comets and~--- more rarely~--- inward transport from the outer belts may be viable, and not mutually excluding, alternatives to the asteroid belt scenario. The remaining five discs (2\% of the sample\revised{: HIP~11486, HIP~23497, HIP~57971, HIP~85790, HIP~89770})
harbour inner components that appear inconsistent with dust production in an ``asteroid belt.''
Warm dust in these systems must either be replenished from cometary sources or represent
an aftermath of a recent rare event, such as a major collision or planetary system instability.
\end{abstract}

\begin{keywords}
planetary systems: protoplanetary discs --
          zodiacal dust --
          minor planets, asteroids, general--
          Kuiper belt: general--
          stars: circumstellar matter
\end{keywords}



\section{Introduction}

Debris discs are belts of comets and asteroids around stars.
Collisions and disintegrations of these bodies release dust.
The thermal emission of this dust in the infrared makes debris discs detectable.

The infrared excess emission can often be approximated by blackbody emission of a certain temperature. Yet a significant fraction of discs requires a second blackbody of a distinctly warmer temperature to be added. A combination of two blackbodies is usually sufficient to reproduce the entire spectral energy distributions (SEDs) from mid-infrared to millimetre wavelengths.
Such two-temperature debris discs were found in recent years to be common
\citep[e.g.,][]{morales-et-al-2011,ballering-et-al-2013,chen-et-al-2014, pawellek-et-al-2014,kennedy-wyatt-2014}.

Performing a careful analysis of the sensitivity limits, \citet{kennedy-wyatt-2014} have shown that there are a number of systems where identification of a warm part of the excess emission on top of the bright stellar photosphere is uncertain. 
They have also demonstrated, however, that the warm component in many systems must be real, although its nature is a matter of debate. One possibility is to attribute the warm excess emission to the dust transported inward from the
main disc by Poynting-Robertson (P-R) effect \citep{burns-et-al-1979,gustafson-1994} or stellar winds \citep{plavchan-et-al-2005}. This has been shown to work for several individual systems \citep[e.g.,][]{reidemeister-et-al-2011,loehne-et-al-2011,schueppler-et-al-2014,schueppler-et-al-2015}. 
The most natural scenario, however, is the ``true'' two-component interpretation.
Many debris disc systems could harbour a Kuiper-belt analogue and an asteroid-belt analogue closer
in \citep[e.g.,][]{su-et-al-2013}.
The presumed inner belt may have been marginally resolved for 
some systems, such as $\varepsilon$~Eri \citep{greaves-et-al-2014} and HD~107146 \citep{ricci-et-al-2015}.

By analogy with the Solar System with its Kuiper belt and asteroid belt,
such a two-component structure could be created by a set of giant 
planets.
If these succeeded to form at the protoplanetary stage, they would carve a hole in the 
protoplanetary disc by removing neighbouring gas, planetesimals, and dust \citep[][and references 
therein]{dipierro-et-al-2016}.
After the gas dispersal, nascent planets would swiftly remove 
planetesimals from their chaotic zones.
At the same time, these planets would dynamically excite planetesimals 
in the zones bracketing
the planetary region, preventing them from growing further to full-size 
planets.
All this would generate a broad gap in the planetary region,
splitting up the disc into two distinct debris belts.

This scenario has already been systematically explored in \citet{schueppler-et-al-2016}.
As a method, they used detailed, but \revised{computationally expensive}, collisional simulations with the code
\texttt{ACE} \citep{krivov-et-al-2006,loehne-et-al-2011,krivov-et-al-2013}.
The application was made to one prominent two-component system, q$^1$~Eridani. Since this kind of analysis is very time consuming even for one system, an extension to hundreds of systems is not feasible. So in this paper, we do a similar analysis, but with an analytic model of
\citet{loehne-et-al-2007}. This allows us to analyse, in a statistical manner,
a large sample of the unresolved debris discs reported to contain two components
\citep{chen-et-al-2014}. The final goal is to find out, whether a majority of the
bona fide two-component debris discs are consistent with the scenario just described and,
if so, which masses and radial profiles the original protoplanetary discs must have had to produce the discs currently seen.

Section~2 describes and refines the debris disc sample.
Section~3 explains the collisional model and the fitting scheme.
The results are presented in Section~4 and discussed in Section~5.
Section~6 draws the conclusions.

\section{Sample}

We used the sample of \citet{chen-et-al-2014}, which encompasses spectra of a total of 571 main-sequence stars with ages up to a few $\Gyr$s. These systems were tested in \cite{chen-et-al-2014} for a two-, one- or zero-blackbody model, and the most probable one was selected through the Bayes theorem. The data points used for the fits were the Spitzer/MIPS 24{\micron} and 70\micron, as well as the Spitzer/IRS spectra, weighted lower in comparison to the MIPS data points.
In the end 333 systems were found, where a two-component blackbody model was the most plausible solution. From these we excluded 72 systems as they either reached the lower ($30\K$) or the higher bound ($500\K$) of the temperature fit. 

This leaves us with a large sample of 261 debris discs, handled in a homogeneous way and thus perfectly suitable for a coherent theoretical analysis. Accordingly, we take both the stellar parameters (mass, luminosity, age) and the derived disc parameters (fractional luminosity and temperature of both components) exactly as published in
\citet{chen-et-al-2014}, as a starting point for our study. The only exception was the radii where the following correction was applied. Since the discs were fitted with blackbodies, whereas the grains in debris discs are not perfect absorbers and emitters, the radii inferred from the blackbody temperature are smaller than the actual radii \citep{booth-et-al-2013,pawellek-et-al-2014}. Analyzing a set of resolved debris discs for which the true radii can be measured directly from the images, \citet{pawellek-krivov-2015} found the ratio of the true radius and the blackbody radius, $\Gamma$, to be a function of the stellar luminosity:

\be
\Gamma =5.42 \left(\frac{L_{\star}}{\revised{\Lsun}}\right)^{-0.35}.
\label{eq:Gamma}
\ee
The coefficients in Eq.~(\ref{eq:Gamma}) \revised{were derived assuming compact} dust grains composed of astrosilicate \citep{draine-2003} and ice \citep{li-greenberg-1998} in equal volume fractions. Since the dust composition is unknown, this choice is rather arbitrary; however, the results would not differ much for silicate particles with moderate porosity or \revised{with} carbonaceous inclusions. In what follows, we use the corrected radii of the cold and warm components, i.e. the blackbody radii calculated from the temperatures listed by \citep{chen-et-al-2014} after multiplication with the $\Gamma$-factor.
This comes with the caveat that the radius correction has never been tested, and may not hold, for the warmer components if they exhibit, for instance, a distinctly different distribution of grain sizes \citep{mittal-et-al-2015}.

The next question to address is how certain the identification of the second component is. We have already discarded some systems, whose warm and cold components appear unrealistically hot and cold, respectively.
However, we have not yet checked the sample for the cases where either component is 
too faint compared to the other, or where both components have similar temperatures. 
Such cases could cast doubt on the two-component interpretation of a disc's SED. To 
address such questions, \citet{kennedy-wyatt-2014} introduced $R_T \equiv T_\text{w}/T_\text{c}$ 
and $R_f \equiv f_\text{w}/f_\text{c}$, the ratios of temperatures and fractional luminosities of 
the warm (``w'') and cold (``c'') components, and used them to differentiate unambiguous two-component discs from probably one-component systems as follows. One assumes a typical central star and a representative cold component, and takes hypothetical warm components with various $R_T$ and $R_f$. For each of the resulting two-component systems, one calculates the synthetic SED and assigns the error bars by assuming typical measurement uncertainties of those instruments that usually deliver such data at the wavelengths involved. Finally, one runs a standard SED fitting routine, trying to fit a one-component model to that SED. The $\chi^2$-value for the fit can serve as a measure of how well the presence of the warm component can be inferred from the measurements. We choose $\chi^2=5$ to distinguish between reliable and unreliable cases.  In other words, $\chi^2>5$ is treated as a failure of a one-component model and is interpreted as a clear indication that the second, warm component is real.

\begin{figure}
\centering
\includegraphics[width=0.48\textwidth, angle=0]{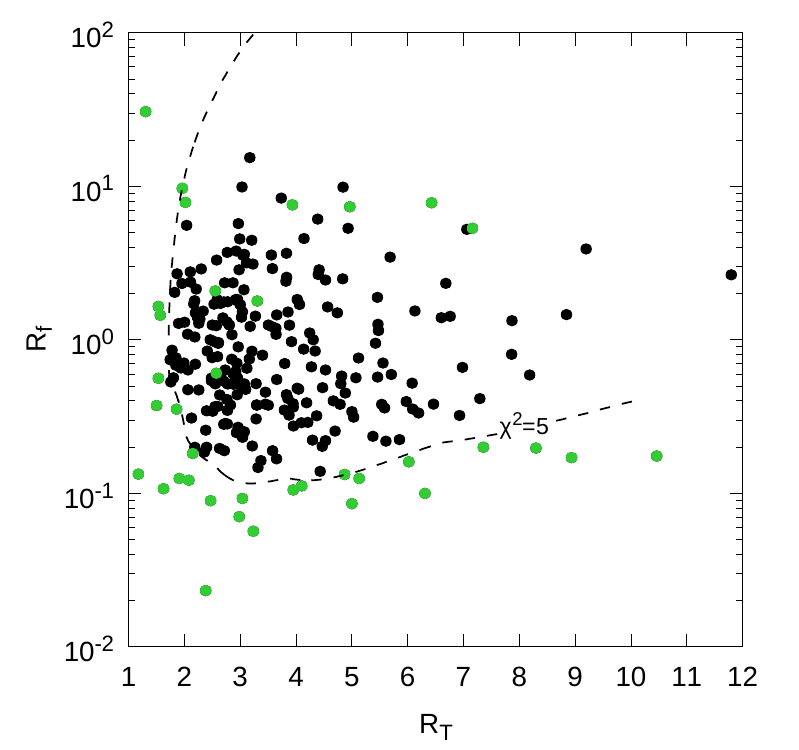}\\
\caption{$R_f=f_\text{w}/f_\text{c}$ plotted over $R_T = T_\text{w}/T_\text{c}$ for the entire sample. The curve implements the method of \citet{kennedy-wyatt-2014}, as explained in the text. To produce this curve, we assumed a star with $L = 9\revised{\Lsun}$ (the log-averaged value for the sample), a cold component at $r_\text{c}=100\AU$ with $f_\text{c}\approx 10^{-4}$, and simulated a grid of warm components with various $r_\text{w}$ and $f_\text{w}$. For each of the resulting fiducial two-component systems, we simulated measurements at 13 wavelengths from near-infrared through sub-mm with a $5\%$ error and \revised{fitted a one-component} model to these measurements. The $\chi^2$ value shown is the reduced $\chi^2$ value attained from these fits. Values of $\chi^2 < 5$ imply that a system is well described by one blackbody;
the cases with $\chi^2 \ge 5$ suggest that a second component is needed.
All systems with $\chi^2 \ge 5$, as well as faint systems with $f_\text{c} < 4 \times 10^{-6}$, are considered unreliable. These are shown in green. \changed{Figure slightly reworked to better use space available.}
}
\label{fig:KWy14plot}
\end{figure}

\begin{figure}
\centering
\includegraphics[width=0.48\textwidth, angle=0]{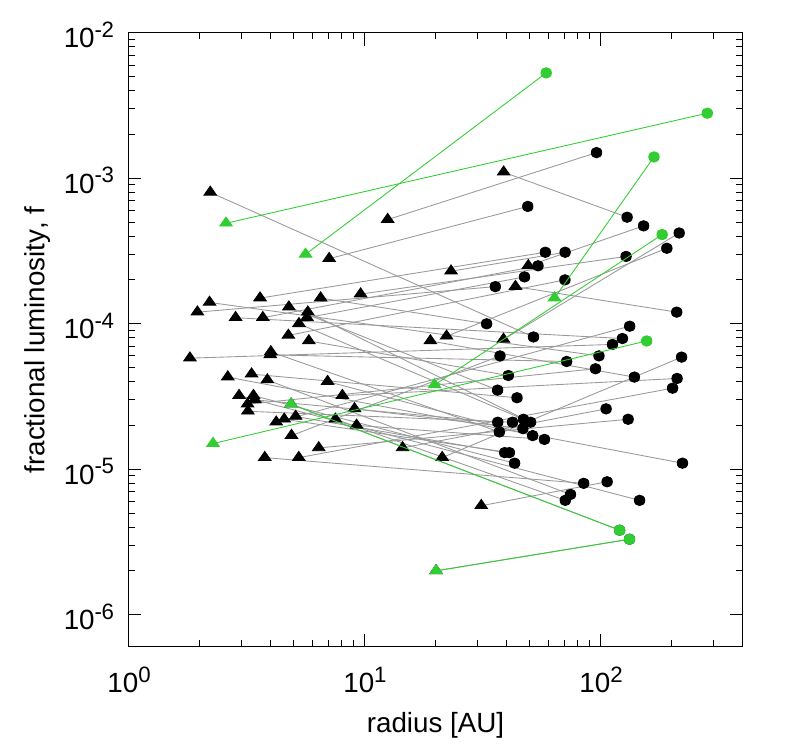}\\
\caption{Two-component systems of our sample on the radius--fractional luminosity plane. Each system is depicted with two symbols which represent the cold (circles) and warm (triangles) components and are connected with a straight line. Black are reliable systems ($\chi^2>5$ and $f_\text{c} > 4 \times 10^{-6}$) while green are unreliable ones ($\chi^2<5$ or $f_\text{c} < 4 \times 10^{-6}$). For better visibility, only one-fifth of all systems are shown. \changed{Figure slightly reworked to better use space available.}
}
\label{fig:radi}
\end{figure}

Applying this model to \revised{our sample}, we get the results visualized in Figs.~\ref{fig:KWy14plot} and \ref{fig:radi}. These two figures use \revised{two complementary ways} of plotting the sample. Figs.~\ref{fig:KWy14plot} shows the discs in the $R_f$--$R_T$ plane, whereas Fig.~\ref{fig:radi} depicts them in the $f$--$r$ plane. Of 261 systems, 29 have $\chi^2<5$, and we dropped them from the sample. This leaves us with 232 systems that \revised{could be truly} two-component ones.

Obviously, this still does not ensure that the remaining sample only contains reliable cases. Most notably, the Kennedy \& Wyatt method does not take into account the absolute value of the fractional luminosity. Systems in which both components have similar, but very low fractional luminosities will be classified as reliable. To avoid this, we removed another seven systems, in which the fractional luminosity of the cold component \revised{was} less than $4 \times 10^{-6}$. This value is the average sensitivity limit of the Spitzer/MIPS instrument at 70{\micron} \citep{bryden-et-al-2006}. Our final sample contains 225 systems that we consider to be reliable two-component discs. These are plotted in Figs.~\ref{fig:KWy14plot} and \ref{fig:radi} in black.

\section{Model}

\subsection{Idea}

\begin{figure}
\centering
\includegraphics[width=0.48\textwidth, angle=0]{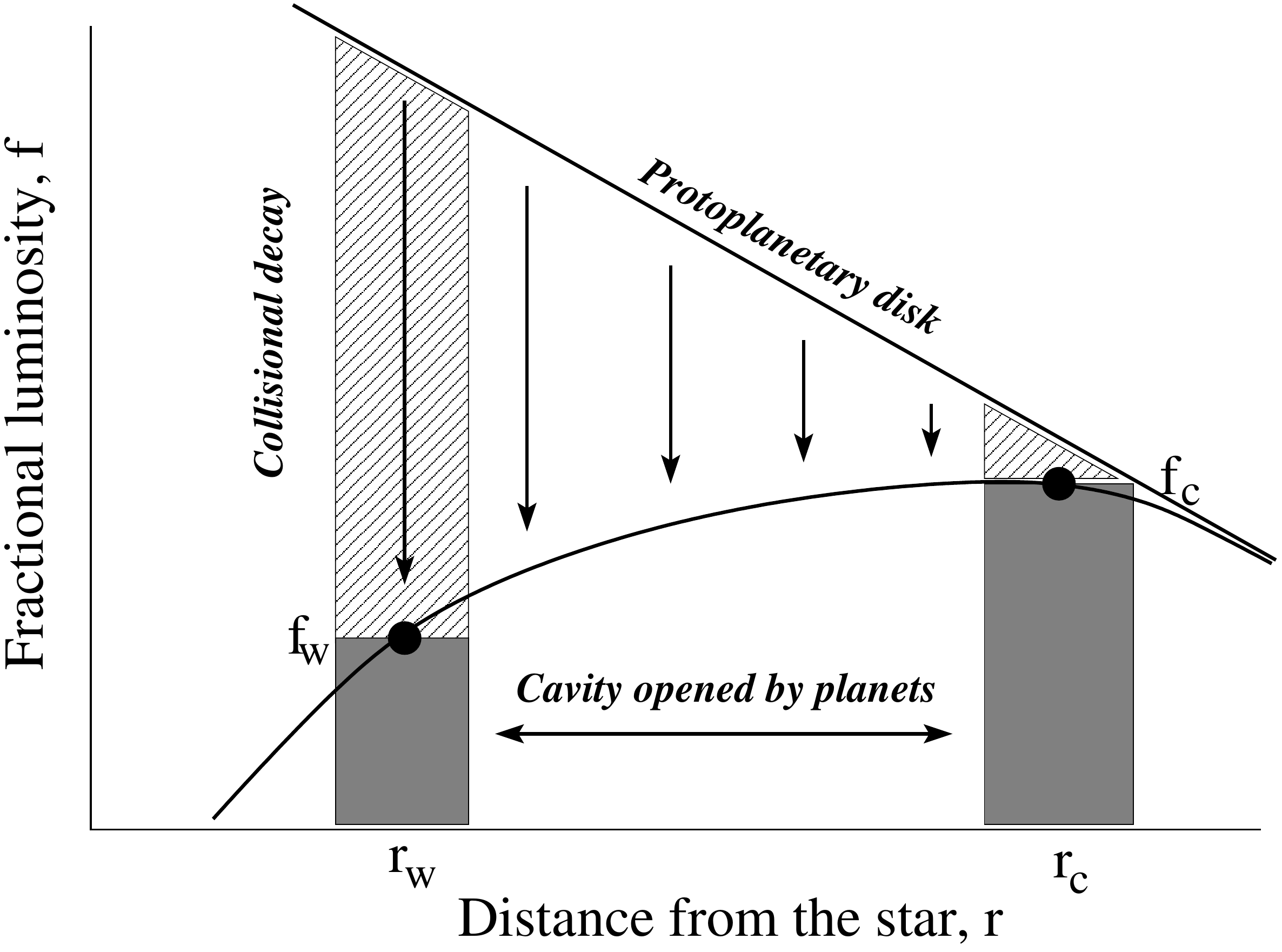}\\
\caption{
Formation of a two-component debris disc.
\label{fig:profit}
}
\end{figure}

The idea of this study is simple (Fig.~\ref{fig:profit}). Each two-component debris disc in the sample is described by two narrow rings of a certain relative width $\total r/r$, the cold one at a distance $r_\text{c}$ with a dust fractional luminosity $f_\text{c}$, and the warm one at a distance $r_\text{w}$ with a dust fractional luminosity $f_\text{w}$. These two pairs of values, $(r_\text{c},f_\text{c})$ and $(r_\text{w},f_\text{w})$, are those depicted in Fig.~\ref{fig:radi}. A natural hypothesis is that this system stems from a protoplanetary disc extending from $r_\text{w}$ to $r_\text{c}$, in which nascent planets (or any other mechanism that we actually do not need to specify) opened a gap between the two rings. After the gas dispersal and ignition of the collisional cascade, these two rings gradually deplete collisionally to reach their observed fractional luminosities $f_\text{w}$ and $f_\text{c}$ by the time $t$, which is the (collisional) age of the system that we set to be equal to the stellar age. We wish to check whether each system is consistent with this scenario and, if so, for which parameters of the original protoplanetary disc.

To prove this hypothesis, we need a collisional model, i.e. a function that describes how the fractional luminosity $f$ of a narrow ring of radius $r$ evolves with time $t$:
\be
  f = f(t, r, \revised{\mathbf{c}}),
\label{f}
\ee
where \revised{$\mathbf{c}$} is a vector of model parameters.
Since we only have the fractional luminosity of the warm and cold component to which we can compare our model, we deliberately have to select no more than two parameters (\revised{$c_1$ and $c_2$}) to vary. The others have to be fixed by making an educated guess; if these are completely unknown, we can afford trying a few distinct choices for such parameters. Formally, this leads to a system of two equations
\be
  f_\text{c} = f(t, r_\text{c}, \revised{c_1, c_2}), \qquad f_\text{w} = f(t, r_\text{w}, \revised{c_1, c_2}),
\ee
to be solved for \revised{$c_1$ and $c_2$}.

These two parameters are chosen as follows. We assume the surface density of solids in the original protoplanetary disc in the form of the ``Minimum-Mass Extrasolar Nebula'' (MMEN; \citeauthor{kuchner-2004} \citeyear{kuchner-2004})
\be
\Sigma (r)= x_\text{scal}\Sigma_0\left(\frac{r}{1\AU}\right)^{p}\label{eq:Sigma0}
\ee
with $\Sigma_0= 1 \revised{\Mearth}/\AU^2$ being the density of the rocks and ices in the Minimum-Mass Solar Nebula (MMSN; \citeauthor{weidenschilling-1977} \citeyear{weidenschilling-1977}; \citeauthor{hayashi-1981} \citeyear{hayashi-1981}). It is this pair of parameters, $x_\text{scal}$ and $p$, which we take as \revised{$c_1$ and $c_2$}. Instead of $x_\text{scal}$, we can also use the total mass of the protoplanetary disc $M_{\text{PPD}}$ that is calculated by integrating Eq.~(\ref{eq:Sigma0}) from $r_\text{w}$ to $r_\text{c}$:
\be
M_\text{PPD}=100 \times 2\pi ~\int_{r_\text{w}}^{r_\text{c}}\Sigma(r)r\total r \label{eq:Mppd},
\ee
where a front factor $100$ is the canonical gas-to-dust ratio \citep{hildebrand-1983}.
To explain the idea in other words, we \revised{seek} a mass and slope of an initial extended disc of solids that would result in a desired amount of material at its inner and outer edges after collisional evolution over the system's age.

We still have to choose the collisional model, i.e., to specify function (\ref{f}). The simplest choice would be a popular analytic model of \citet[][their Eq. 20]{wyatt-et-al-2007}. However, this model may be too crude for our purposes \citep{schueppler-et-al-2016}, and we choose to employ a more detailed model by \citet{loehne-et-al-2007}. It is described in the Appendix. The function $f = f(t, r, \revised{\mathbf{c})}$ of this model is given by Eq.~(\ref{eq:fdt}).

\subsection{Parameters}

We now discuss the parameters of the collisional model (see Table~\ref{tab:param}).

\begin{table}
\caption{
Model parameters.
 \label{tab:param}}\begin{center}
\begin{tabular}{lcc}
\hline
Parameter	& Value of range	& Warm/cold comp.\\
\hline
\multicolumn{3}{c}{\em Known parameters}\\
$L_\star$, $M_\star$	& From Chen+14	& ---\\
Age 			& From Chen+14	& ---\\
$f_\text{c}$, $f_\text{w}$ 		& From Chen+14	& ---\\
$r_\text{c}$, $r_\text{w}$		& Calc. from $T_\text{c}$, $T_\text{w}$&\\
                 & from Chen+14, & ---\\
			& multiplied by $\Gamma$	&\\
\hline
\multicolumn{3}{c}{\em Fixed parameters}\\ 
$e$		& 0.01,0.03,0.1,0.2	& Same\\
$\total r/r$	& 0.1, 0.3		& Same or different\\
$s_{\text{max}}$& $\le100\km$		& Same\\
$s_\text{D}$		& $ 1\mm$			& Same\\
$A_\text{s}$, $A_\text{g}$ & $5\times10^6 \erg/\g^{-1}$ & Same\\
$b_\text{s}$		& $-0.12$		& Same\\
$b_\text{g}$		& $ 0.45$		& Same\\
\hline
\multicolumn{3}{c}{\em Free parameters}\\
$p$			& $[-4,1]$	& ---\\
$\log x_\text{scal}$	& $[-4,2]$	& ---\\
\hline
\end{tabular}
\end{center}
\end{table}

{\em Original disc.}
\revised{The original solid density profile was described with two parameters.}
The slope $p$ was allowed to vary from $p = -4$ to $p=1$ covering reasonable values, expected for the protoplanetary disc (with the MMSN having $p=-1.5$). The scaling factor $x_{\text{scal}}$ was probed in the range between $\log x_\text{scal}=-4$ and $\log x_\text{scal}=2$. The \revised{upper} boundary was imposed to exclude extremely massive discs (especially those that would be unstable against self-gravity). The lower boundary is just a reasonable minimum for the protoplanetary disc mass.

{\em Central star.}
The host star is characterized by its luminosity $L_\star$ and mass $M_\star$. Both parameters, together with the dust density and optical properties, determine the radiation pressure blowout limit for dust, which we take as the lower cutoff for \revised{the} sizes of solids $s_\text{min}$. Besides, the stellar mass sets the impact velocities, see Eq.~(\ref{eq:Xc}).

{\em Planetesimal belt.}
One parameter is the relative disc width $\total r/r$, which needs to be specified to \revised{enable} conversion between the surface density $\Sigma$ of material in the extended disc and mass (or fractional luminosity) of both cold and warm components.  For instance, we need this parameter to calculate the initial mass of each component that enters Eq.~(\ref{eq:fdt}). We tried $\total r/r$ of 0.3 and 0.1, allowing \revised{the components to have unequal values.} Another parameter is the dynamical excitation level of a planetesimal ring, parameterized by eccentricities $e$ and inclinations $I$. To reduce the number of parameters, we eliminated inclinations by assuming equipartition condition $e=I/2$. Similarly to the disc width, we \revised{considered} several values of eccentricity between 0.01 and 0.2, \revised{setting the same} value for both components. We excluded combinations of $\total r/r$ and $e$ such that $e > \total r/r$ to ensure that the radial excursions of particles moving in elliptic orbits do not not make a disc wider than $\total r$. Similarly, we checked $r_\text{c}$ and $r_\text{w}$ against $\total r/r$ to exclude overlapping discs with no gap between the components.

{\em Mechanical properties of solids.}
Apart from the bulk density $\rho$, which we arbitrarily set to $2700\kg/\m^3$, essential parameters are 
those describing the critical energy needed to disrupt the solids of different sizes.
Small particles are in the strength-dominated regime (``s''), while
large objects are in the self-gravity-dominated regime (``g'').
The size-dependent critical energy for fragmentation is then represented by a sum of
two power laws of the object's size, with coefficients
$A_\text{s}$ and $A_\text{g}$ and exponents $b_\text{s}$ and $b_\text{g}$.
Their assumed values are given in Table~\ref{tab:param}.
See Eq.~(\ref{eq:QD}) in the Appendix for more explanations.

{\em Characteristic sizes of solids.}
The two parameters $s_\text{min}$ and $s_\text{D}$ determine the minimum and maximum radii of particles in the discs that 
we refer to as ``dust.'' The former is set to the radiation pressure blowout limit \citep{burns-et-al-1979} 
and is on the order of $\sim 1$\micron. The latter can be set to an arbitrary value that is large enough to 
ensure that material with $s>s_\text{D}$ \revised{makes} a negligible contribution to the cross section (or fractional luminosity); we take
$s_\text{D} = 1\mm$.
\revised{Another parameter} that affects the collisional evolution significantly is the radius of the largest planetesimals in the disc $s_{\text{max}}$. It was initially set to $100\km$ for both components. Yet at greater distances from the star the relative velocities of the objects reduce. Hence large objects are harder to destroy in a collision. With increasing radius the minimum size of a destructive projectile given by Eq.~(\ref{eq:Xc}) increases. If $X_\text{c}(s_\text{max})$ becomes larger than unity,  planetesimals of the size $s_\text{max}$ could only be destroyed by larger objects, which are absent.
To avoid this, we require that $X_\text{c}(s_\text{max}) \le 0.75$ and reduce the
 $s_{\text{max}}$ accordingly. This means objects larger than our $s_\text{max}$ exist in the discs, but do not take part in the collisional cascade and are not considered.

\begin{figure*}
\centering
\includegraphics[width=0.98\textwidth, angle=0]{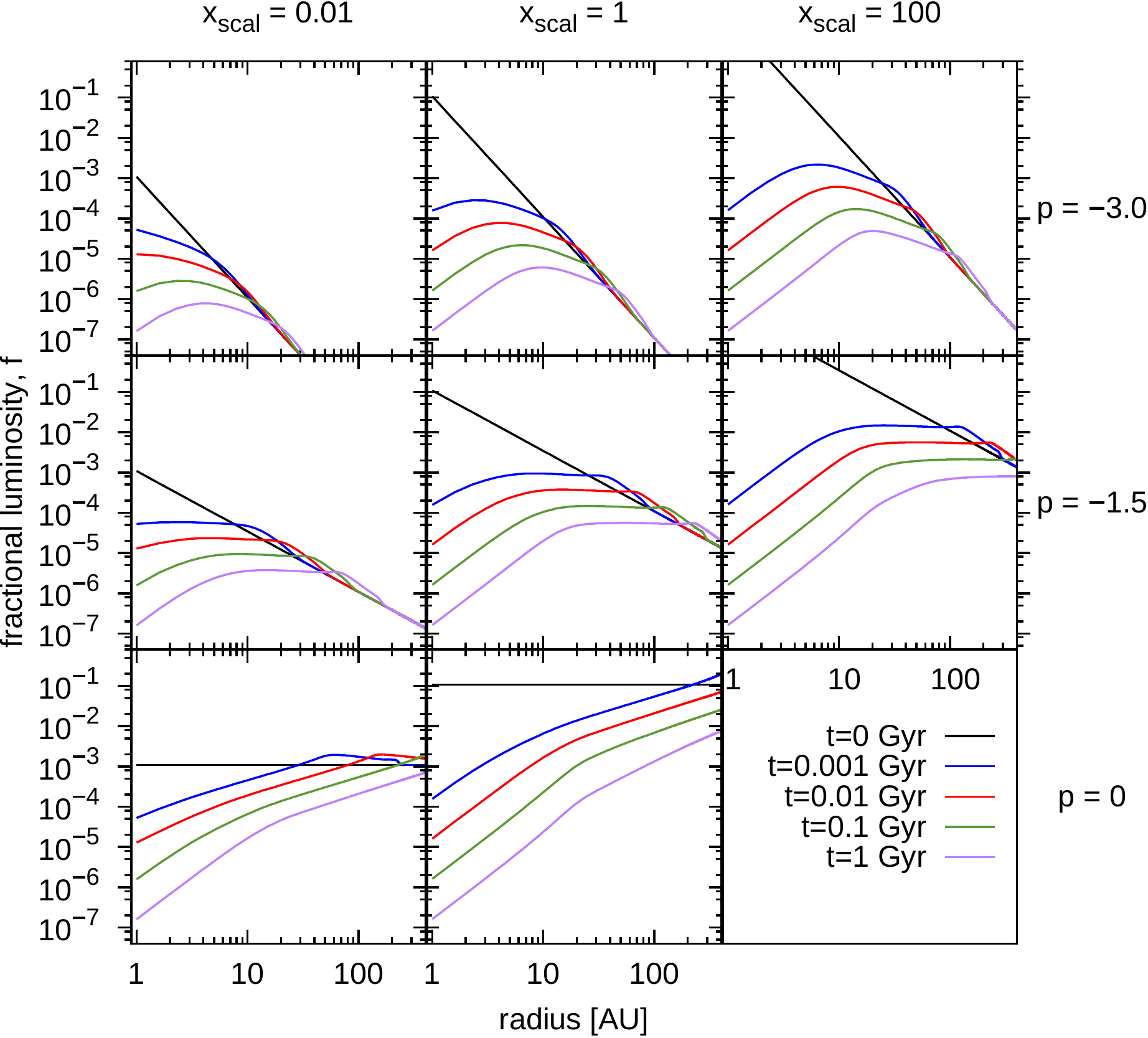}\\
\caption{
Radial profiles of the fractional luminosity in fiducial discs of various initial masses and slopes.
From left to right: $x_\text{scal} = 0.01$, 1, and 100.
\revised{From} top to bottom: $p=-3$, -1.5, and 0.
The solid lines are profiles calculated with the \citet{loehne-et-al-2007} model.
The central star a luminosity of $9\revised{\Lsun}$ and a mass of $1.8\revised{\Msun}$ was used (both being the log-averaged values for the sample). The disc parameters in both models were $\total r/r=0.1$, $e=0.1$ and $s_{\text{max}}=100\km$. \revised{Where necessary, $s_{\text{max}}$ was reduced} at larger radii, to allow the largest objects to still be destroyed by smaller ones and thus to be part of the collisional cascade. 
\label{fig:9panels}
}
\end{figure*}

{\em Slopes of the size distribution.}
The model by \citet{loehne-et-al-2007} works with a three-slope size distribution. For solids in the strength regime, the slope is $q_\text{s}$. For objects that are large enough to reside in the gravity regime, but small 
enough to be in a collisional equilibrium at the time instant considered, the slope is $q_\text{g}$. Finally, the 
largest planetesimals that are not (yet) involved in the cascade, since their collisional lifetimes are 
longer than the current age of the system, retain their primordial size distribution with a slope $q_\text{p}$. Of 
these three slopes, the first two are calculated from $b_\text{s}$ and $b_\text{g}$ 
\citep{durda-dermott-1997,o'brien-greenberg-2003}, as explained in the Appendix. This leads to 
$q_{\text{s}}=1.89$ and $q_{\text{g}}=1.68$. The third one has to be postulated, and we take $q_\text{p}=1.87$, 
following \citet{loehne-et-al-2007}.

\subsection{Examples}

\revised{The model} described in the previous subsections \revised{can be used to} illustrate the evolution of fiducial 
discs with different initial mass ($x_\text{scal}=0.01$, 1, and 100) and radial profiles of surface density ($p=-3$, -1.5, 0). We consider a central star of $1.8\revised{\Msun}$ and $9\revised{\Lsun}$. The modelled disc was assumed to have a width of $\total r/r=0.1$ and planetesimal eccentricity of $0.1$. 

The evolution calculated with the model of \citet{loehne-et-al-2007} is shown in Fig.~\ref{fig:9panels}. To discuss it, we move from the 
outer regions of Fig.~\ref{fig:9panels} inward. Since the collisional evolution far from the star is the slowest, the outermost regions in the plot are still primordially distributed. Closer in, the model predicts the fractional luminosity to rise: the coloured lines are slightly higher than the primordial distribution. 
\changed{Three sentences deleted.} \revised{This is because the smallest grains enter the cascade first} and then develop a steeper size distribution than the primordial one ($q_\text{s} > q_\text{g}$). This leads to more abundant small grains and therefore to \revised{a rise} in the radial fractional luminosity profile above the primordial values as seen in Fig.~\ref{fig:9panels}.

Still further inward, the fractional luminosity drops below the primordial one. This occurs because the initial \revised{pile-up} of smaller dust grains comes to a halt when the objects entering the collisional cascade become large enough \revised{to reach the self-gravity regime}. These objects follow a slope shallower than the primordial one ($q_\text{g} < q_\text{p}$), \revised{causing the fractional luminosity to drop}. \changed{Three sentences deleted.}This stage of the evolution persists up to the point at which the collisional lifetime of the largest objects is reached. These start to deplete, as no larger objects can replenish them through collisions, and the fractional luminosity reduces \revised{more rapidly}.

Closest to the star, the disc evolution has progressed to even later stages and the depletion has been in effect for a long time. At these stages the fractional luminosity of \citet{loehne-et-al-2007} becomes independent \revised{of} the initial mass of the disk and \revised{takes} similar values at similar radii regardless of $x_\text{scal}$. This is best visible in the right column of Fig.~\ref{fig:9panels} that depicts most massive discs. In the left column, i.e., for the lightest discs, that region is located outside the left edge of the plot and is not seen.

\revised{Prior to fitting the collisional model into the sample of discs, we may directly compare the model predictions (Fig.~\ref{fig:9panels}) with the actual sample (Fig.~\ref{fig:radi}). Both figures use the same axes and similar plotting ranges. We see, for instance, that systems with $f_\text{w} \ll f_\text{c}$ (i.e., with warm components fainter than the cold ones) have good chances to be reproduced by models, unless both components have low luminosity. Also, it should be easier to find a matching model for younger systems. All this is a conseqence of the fact that collisional depletion occurs faster in more massive belts, and that belts located closer to the star decay faster than more distant ones.
}

\section{Results}
We \revised{applied the fitting procedure to our sample, searching} for the best solutions for ($p$, $x_\text{scal}$) over all possible combinations of $\total r/r$ \revised{and $e$}. With all possible sets of parameters we found a good fit for 220 of the 225 examined systems.
We call a fit ``good'', if the relative deviations of the modelled fractional luminosities from the observed ones added in quadrature are smaller than $10\%$.

\begin{figure}
\centering
\includegraphics[width=0.48\textwidth, angle=0]{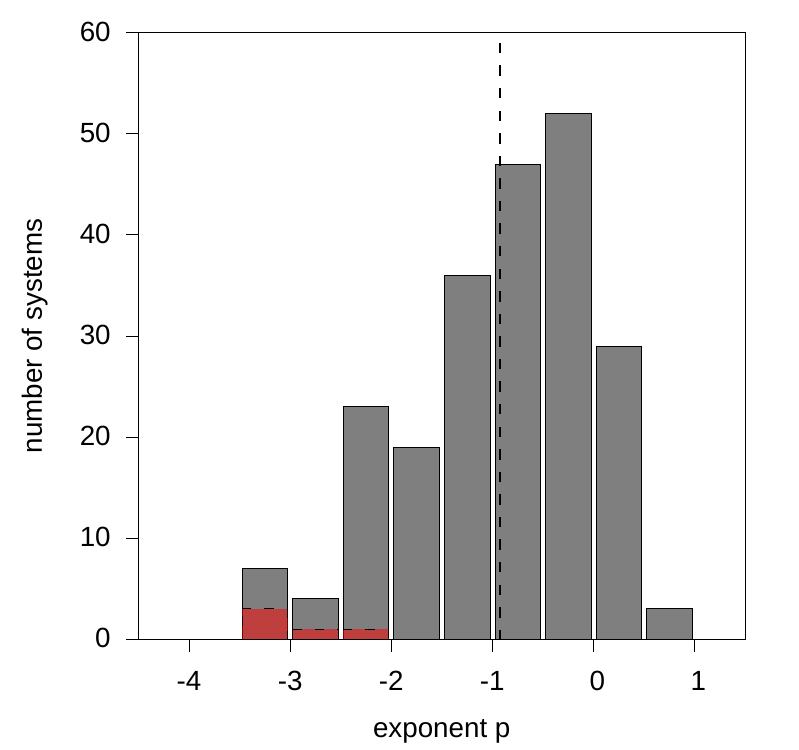}\\
\caption{Distribution of radial profile slopes $p$. Black: all systems with good fits, red: unsuccessful fits.
}
\label{fig:histograms_p}
\end{figure}

\begin{figure}
\centering
\includegraphics[width=0.48\textwidth, angle=0]{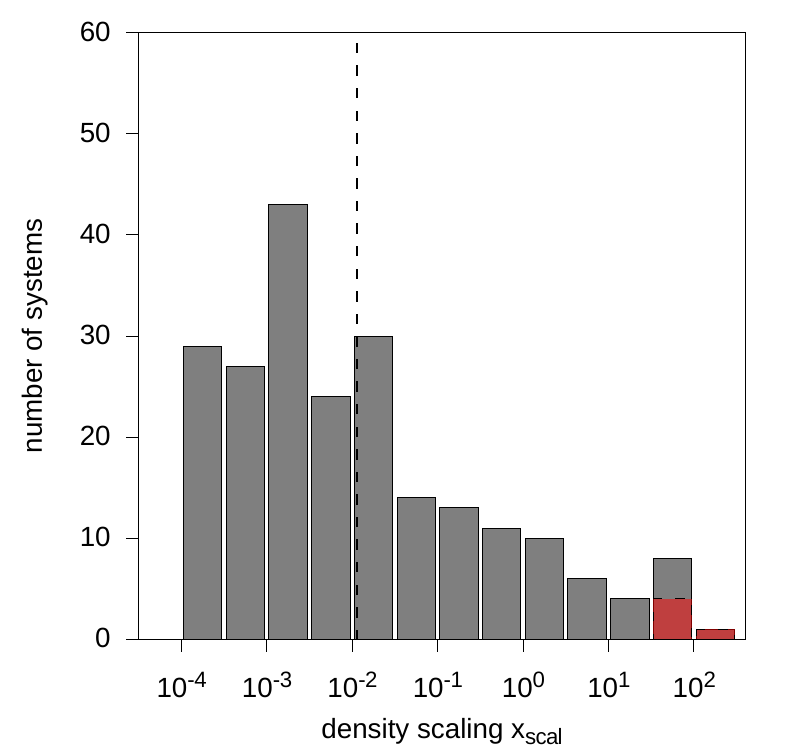}\\
\caption{Distribution of scaling factors $x_\text{scal}$. Colours are the same as in Fig.~\ref{fig:histograms_p}.
}
\label{fig:histograms_x}
\end{figure}

In Fig.~\ref{fig:histograms_p} the distribution of exponents $p$ of successful fits is depicted in black and bad fits in red. The distribution is asymmetric with \revised{a sharp} drop off for values greater than $0.5$.
The mean value of the exponent is $-0.93\pm0.06$. This is rather far from the MMSN slope $r=-1.5$. It is also different from the MMEN slope of $p \approx -1.6$ derived from the analysis of systems with super-Earth-type planets \citep{chiang-laughlin-2013}. Nevertheless, it is consistent with the slopes of protoplanetary discs inferred from submillimetre observations \citep[][and references therein]{williams-cieza-2011}.
\changed{Sentence deleted.}
The fact that the distribution is broad may stem from the observational uncertainties, as well as from the imperfectness of our approach. However, it may also reflect the real dispersion of the profiles in protoplanetary discs.  For instance, \citet{raymond-cossou-2014} argue against a ``universal slope'' and derive a distribution of exponents in the range $-3.2...0.5$ by constructing the MMEN profiles in systems with multiple low-mass planets.  

The density scaling distribution is plotted in Fig.~\ref{fig:histograms_x}. The mean value is $x_\text{scal} =10^{-1.94\pm0.10}$, but the distribution is broad and highly asymmetric, with more systems exhibiting smaller $x_\text{scal}$. Since the scaling factor is ``anchored'' to $1\AU$, i.e. to the innermost region of the disc, this implies tenuous warm discs that are too faint in comparison to the model.
And conversely,  systems with large $x_\text{scal}$ indicate that the warm component is brighter than what is expected from the model. All systems with poor fits belong to this category.

\begin{figure}
\centering
\includegraphics[width=0.48\textwidth, angle=0]{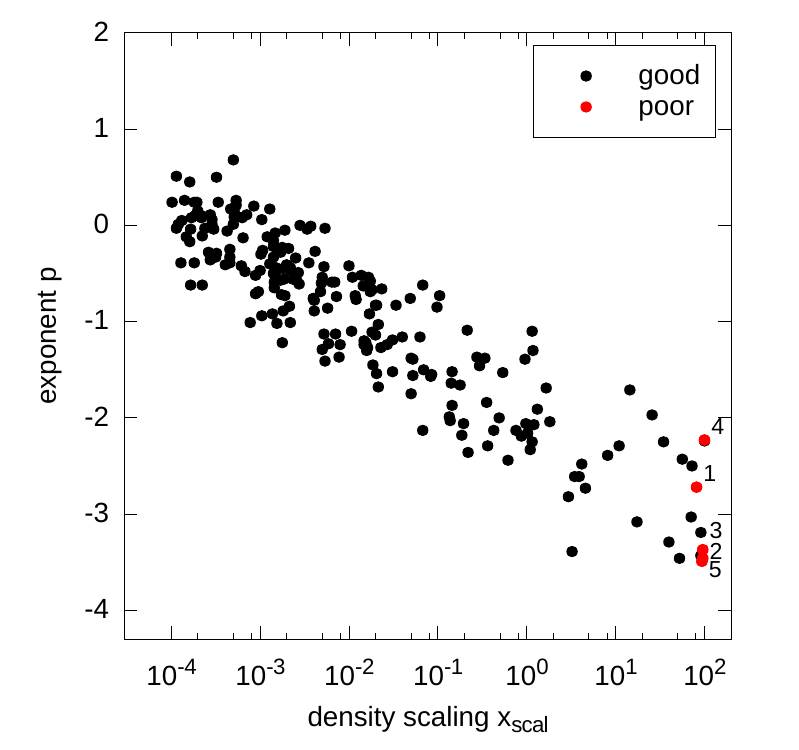}\\
\caption{Relation between the density scaling factor $x_\text{scal}$ and exponent $p$. The black symbols stand for the well fitted systems and the red circles for the poorly fitted ones.
All systems with poor fits are labelled and listed in Table~\ref{tab:systems}.
\changed{Sentence deleted.}
\label{fig:pvsscal}
}
\end{figure}

\begin{figure}
\centering
\includegraphics[width=0.48\textwidth, angle=0]{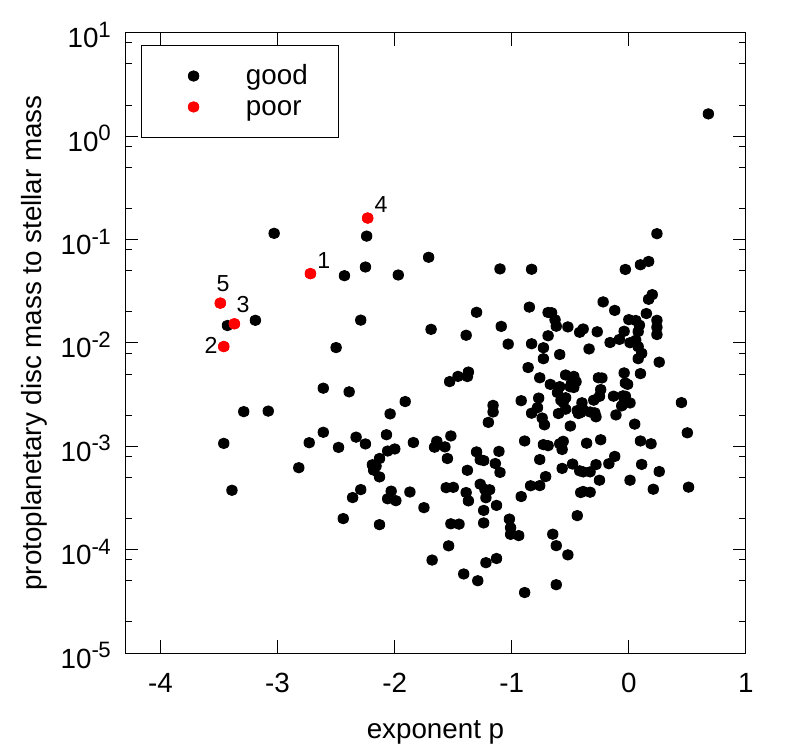}\\
\caption{Relation between the total disc mass to stellar mass ratio $M_\text{PPD}/M_\star$ and exponent $p$.
\changed{Sentence deleted.}
Symbols are as in Fig.~\ref{fig:pvsscal}. An outlier at the top right with an unrealistically large disc mass
is a young ($\sim 5\Myr$) system HD~36444 that may be a protoplanetary or transitional, rather than debris, disc \citep{hernandez-et-al-2006}.
\label{fig:pvsmass}
}
\end{figure}

\begin{figure}
\centering
\includegraphics[width=0.48\textwidth, angle=0]{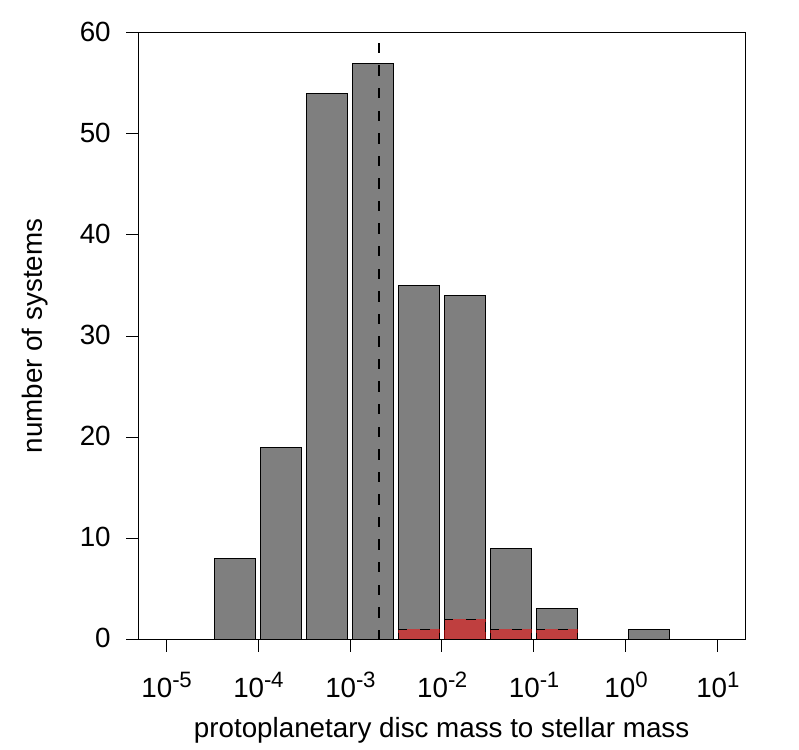}\\
\caption{Distribution of the initial masses of the protoplanetary disc progenitors, in the units of stellar mass. The mass of solids was multiplied with the gas to dust ratio of $100:1$.
\label{fig:histograms_m}
}
\end{figure}

In fact, the parameter $x_\text{scal}$ alone is not a direct measure of the disc mass. 
The same disc mass may correspond to a lower density at $1\AU$ (i.e., a smaller $x_\text{scal}$) in 
combination with a flatter surface density profile (i.e., a larger $p$)~-- or to a larger
$x_\text{scal}$ along with a smaller $p$.
Figure~\ref{fig:pvsscal} demonstrates that $x_\text{scal}$ and $p$ are indeed strongly correlated, with lower scaling factors corresponding to flatter slopes and vice versa.
To cope with this problem, we can analyze the initial mass of the protoplanetary disc $M_\text{PPD}$ 
directly instead of $x_\text{scal}$. 
The mass was calculated by means of Eq.~(\ref{eq:Mppd}).
Figure~\ref{fig:pvsmass} demonstrates that, unlike the scaling factor, the disc mass is nearly uncorrelated with the slope.

The distribution of disc masses, measured in the units of stellar mass, is shown in Fig.~\ref{fig:histograms_m}. It is nearly symmetric and is centered at
$M_\text{PPD}/M_\star = \left(2.0^{+0.3}_{-0.2}\right)\times 10^{-3}$.
Measuring the disc masses in solar masses, instead of taking the mass of the respective central star as a unit, leads to a similar distribution. In that case, the mean disc mass is $M_\text{PPD}/\revised{\Msun}=\left(3.3^{+0.4}_{-0.3}\right)\times 10^{-3}$.
These results are in good agreement with the mass of protoplanetary discs derived from submillimetre observations. For example, \citet{williams-cieza-2011} report most of the discs to have
$M_\text{PPD}/M_{\star}$ between $10^{-3}$ and $10^{-2}$.

We have also looked for a possible correlation between $M_\text{PPD}$ and $M_{\star}$. We only found a subtle trend of protoplanetary discs being more massive around more massive primaries. However, the scatter of data is large and this trend is statistically insignificant. Thus we can neither confirm nor rule out an approximate scaling $M_\text{PPD} \propto M_{\star}$, which was derived previously from surveys spanning much broader ranges of stellar masses \citep[see][and references therein]{williams-cieza-2011}.

\section{Discussion}

\revised{In the previous section, we fitted the models to a sample of debris discs to infer key parameters of their progenitors, protoplanetary discs. The masses and density slopes of these progenitors obtained in these fits are not in any way unique. Varying different parameters of the model, e.g. materials, eccentricity and initial distributions, would alter the parameters of original protoplanetary discs we deduce.}
Nevertheless, the results allow us to distinguish the systems that can easily be described by a steady-state collisional cascade from those where it fails to reproduce the measurements (at least under the assumptions we made), suggesting that other mechanisms might be at work.

\begin{figure}
\centering
\includegraphics[width=0.48\textwidth, angle=0]{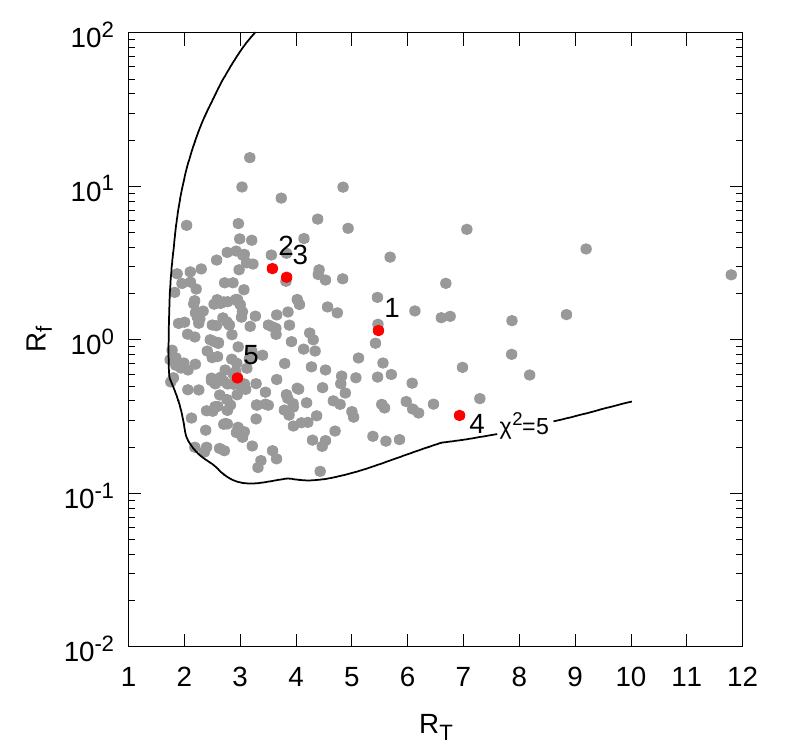}\\
\caption{Same as Fig.~\ref{fig:KWy14plot}, except that grey circles show systems with good fits and red circles depict systems with poor fits. All poorly fitted systems are labelled and listed in Tab.~\ref{tab:systems}. \changed{Figure slightly reworked to better use space available.}
}

\label{fig:KWy14plot2}
\end{figure}

\begin{figure}
\centering
\includegraphics[width=0.48\textwidth, angle=0]{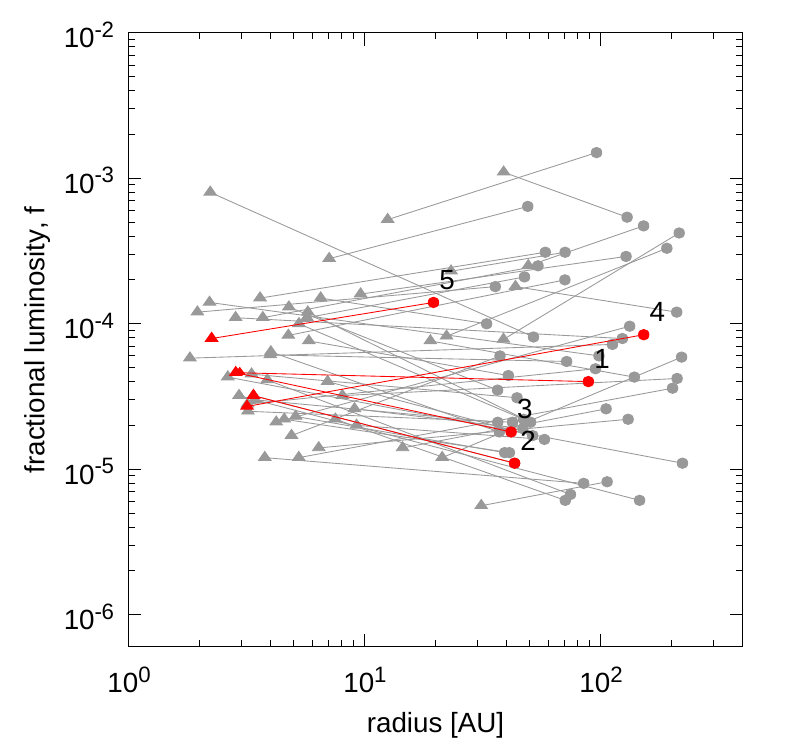}\\
\caption{Same as Fig.~\ref{fig:radi}, except that grey symbols stand for systems with good fits and red symbols denote systems with poor fits. For better visibility only one-fifth of all systems are shown. \changed{Figure slightly reworked to better use space available.}
}
\label{fig:radii}
\end{figure}

To illustrate this, we start with Figs.~\ref{fig:KWy14plot2} and~\ref{fig:radii}. These visualize our sample in the same way as Figs.~\ref{fig:KWy14plot} and~\ref{fig:radi}, except that they \revised{contrast
the systems with good fits (in grey) to} the systems with poor fits (in red).
The overwhelming majority of the systems (220/225) are compatible with a two-component disc, the both components of which have undergone collisional depletion in a steady-state regime. However, there are also five systems that are not compatible with this scenario.
These two classes of systems are discussed in turn in the next two subsections.

\subsection{Systems with good fits}

\begin{figure}
\centering
\includegraphics[width=0.48\textwidth, angle=0]{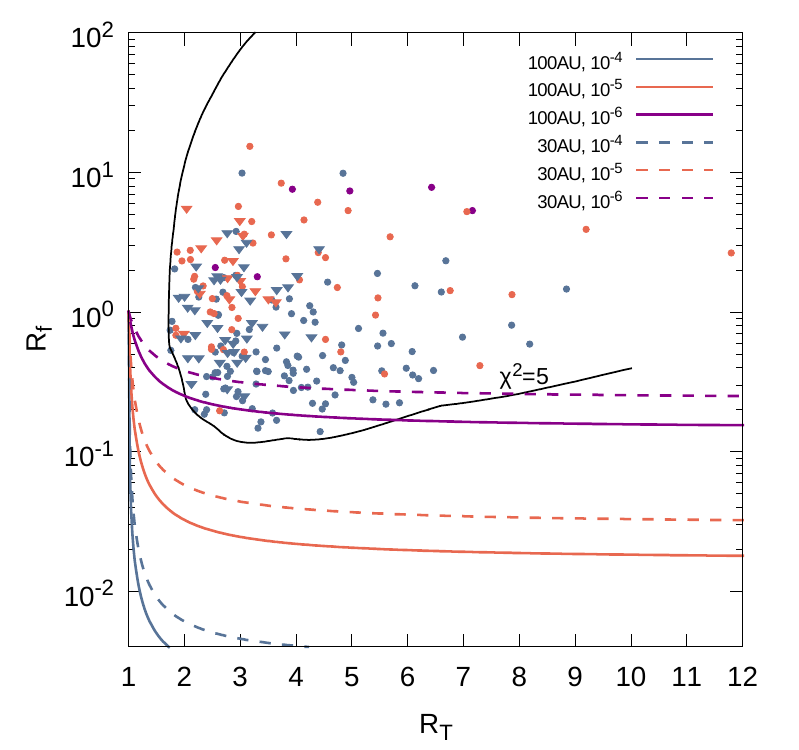}\\
\caption{Same as Fig.~\ref{fig:KWy14plot2}, but containing systems with good fits only.
Curves are loci of several fiducial discs, in which dust is produced in outer parent belts and is transported inward by \hbox{P-R} drag, experiencing collisions \citep{kennedy-piette-2015}.
The central star has a mass $1.8\revised{\Msun}$, each component has a relative width of $\total r/r=0.1$.
Different linestyles correspond to different radii of the outer components 
(solid: $100\AU$, dashed: $30\AU$), whereas different curve colours reflect different fractional luminosities of the outer components (magenta: $10^{-6}$, orange: $10^{-5}$, blue: $10^{-4}$).
Similarly, different symbol shapes characterize the disc size (circles: $r_\text{c}>50\AU$, triangles: $r_\text{c}<50\AU$), while symbol colours refer to their fractional luminosity (magenta: $\sim 10^{-6}$, orange: $\sim 10^{-5}$, blue: $\sim 10^{-4}$). A disc depicted with a circle/triangle of a certain color is explainable with the transport scenario, if it is located below the solid/dashed line of the same color.
}
\label{fig:ParaPlots}
\end{figure}

\revised{Although these} systems are compatible with the asteroid belt \revised{scenario, this is not} the only way of interpreting them. One alternative is that the observed warm dust is transported inward from the outer belt (see Sect.~1.). Pure transport without subsequent collisions would lead to a flat optical depth profile, independent of the distance from the star \citep[e.g.,][]{briggs-1962}. However, the dust experiences collisions on its way inward, and gets progressively depleted, which forces the optical depth to decrease closer to the star \citep[e.g.][]{wyatt-2005}.
With the \revised{decreasing optical depth}, the collisions become less important, which flattens the
optical depth profile again. To describe this interplay of transport and collisions, we used the analytic model of \citet{kennedy-piette-2015}. The normal geometrical optical depth given by their Eqs.~(1)--(2) was converted to the fractional luminosity via $f = (1/2) (\total r/r) \tau$, where we assumed $\total r/r=0.1$. This allows us to ``map'' the discs that would be created by transport and are subject to collisions onto the $R_f$--$R_T$ plane. We do this for fiducial cold belts of two typical radii ($r_\text{c}=100\AU$ and $30\AU$) and three characteristic fractional luminosities ($f_\text{c} = 10^{-4}$, $10^{-5}$, and $10^{-6}$) in Fig.~\ref{fig:ParaPlots}. The ``low collisional rate'' as defined in \citet{kennedy-piette-2015} was assumed.

Comparing the curves with the sample, we conclude that only a few (4/220, or 2\%) of the discs can be created by the transport of dust grains from the outer belt. Such systems usually contain relatively \revised{faint and compact cold} components, and even fainter warm ones. A caveat is that the \citet{kennedy-piette-2015} model is only valid for discs around stars of at least a solar luminosity. The reason is that it assumes drag to be caused by the P-R force only.
However, for late-type stars of subsolar luminosities the pure P-R drag can be drastically enhanced by strong stellar winds \citep{plavchan-et-al-2005}.
This may reinforce the transport scenario as a possible explanation of the observed warm dust in discs of late-type stars  \citep[see, e.g.,][]{reidemeister-et-al-2011, schueppler-et-al-2014,schueppler-et-al-2015}. Since, however, only a minority of stars in the sample have late spectral types, the fraction of systems that can potentially be explained by transport would increase only slightly.  In other words, transport is a plausible alternative to local collisional production in ``asteroid belts'' for some systems, but not the majority of them. 

Apart from asteroid belts and transport from the Kuiper belts, other mechanisms are also possible. For instance, warm dust can be supplied by comets. From the modeling that has been done here, it is difficult to distinguish between the asteroidal and cometary scenario, as both imply dust production in-situ, i.e. they consider dust to be produced where it is observed.

It is interesting to look at our Solar System's debris disc to see what
our analysis would show, and how this compares to the mechanisms of dust production that are fortunately known in this case. The disc in the Solar System is known to be so tenuous that its analogues around other \textcolor{red}{stars} could not yet be detected with current instruments \citep[e.g.,][]{booth-et-al-2009,greaves-wyatt-2010,vitense-et-al-2012}. However, were it detectable, it would also reveal a two-component structure \citep[e.g.,][]{nesvorny-et-al-2010}. The infrared excess in the SED would be comprised of two components~-- the one from the dust residing in the outer system outside Neptune's orbit and another one inside Jupiter's orbit (called the zodiacal cloud). Both have fractional luminosities of $\sim 10^{-7}$ \citep{vitense-et-al-2012,roberge-et-al-2012}. These are so low that collisions are unimportant at dust sizes. Therefore, the fitting procedure employed in the paper would find the system to be consistent with the asteroid belt scenario. Indeed, some of the zodi dust is produced via collisions in the asteroid belt. However, P-R transport also plays a crucial role in shaping the inner dust cloud. \revised{Furthermore, the} prime source of the zodi dust is Jupiter-family comets rather than asteroids \citep{nesvorny-et-al-2010}. Those comets split and disintegrate, leaving trails of \revised{meteoroids that gradually spiral toward the Sun, producing} observable dust in mutual collisions. 
Thus it is a combination of several scenarios, rather than a single one, that works in the Solar
system case. The same complex picture can obviously be true for debris discs of other stars.

\subsection{Systems with poor fits}

We now turn to a discussion of five systems with poor fits.
All these appear on the \revised{right in Fig.~\ref{fig:pvsscal}}, suggesting a high density of the original disc at 1~AU. This is indicative of inner components that are atypically bright for the systems' age. The model has difficulties to retain enough material in the inner region against collisional depletion and tries to counteract this by taking very high initial densities, i.e., a large $x_\text{scal}$. This requires a steep radial slope with $p$ down to $-3.5$, in order to match the observed outer component of ``normal'' brightness (and the fitting routine usually succeeds to reproduce it).

\begin{table}
\caption{
Systems with poor fits.
\label{tab:systems}}\begin{center}
\begin{tabular}{clcrcc}
\hline
No.	& System & \multicolumn{2}{l}{Radius [AU]}& \multicolumn{2}{l}{$f_\text{obs} \times 10^{-5}$}\\
&& warm & cold & warm &cold\\
\hline
1 & HIP~11486 & 3 & 89 & 4.6 & 4.0\\
2 & HIP~23497 & 3 & 43 & 3.2 & 1.1\\
3 & HIP~57971 & 3 & 42 & 4.6 & 1.8\\
4 & \revised{HIP~85790} & 3 & 152 & 2.7 & 8.4\\
5 & HIP~89770 & 2 & 20 & 7.9 & 14\\
\hline
\end{tabular}
\end{center}
\end{table}

For these systems, we can neither reproduce the observed values with the model nor find any reason to doubt the data. There are two logical possibilities. Either our model or parameter choices are unreasonable for these systems~--- or they are truly incompatible with the scenario considered here and warm dust is produced in processes other than a steady-state collisional cascade in an ``asteroid belt.''

Before rejecting the entire scenario, we try to critically re-consider the model and parameter settings. For example, both components may not share the same eccentricity, as we assumed. A way to improve the poor fits would be to reduce the planetesimal eccentricities in the inner component, while keeping or increasing the eccentricities in the outer one. This would lead to an overall lower density with a shallower slope, aligning \revised{it better} with the results for the other systems. To achieve this, we would need a rather strange architecture, with a sufficiently massive planet in an eccentric orbit far out in the system (to increase the eccentricities in the outer component) and a set of closer-in low-mass planets in nearly circular orbits (carving the gap between the two belts, but leaving the inner component nearly unstirred). Furthermore, we checked that the eccentricities in the warm component must be on the order $e \sim 0.001$. Such low values would imply collisional velocities that are too low for collisions to be destructive \citep[e.g.,][]{krivov-et-al-2013}.

Another possibility could be that the dust is made of different materials at different distances from the star. For example we did not consider the ice line in each system while fitting. This would change the composition of the outer components by adding ice to the mixture, influencing the evolution. \revised{Nevertheless, since} the inner components are generally too warm to harbour ice ($ \ga 150\K$), this would not change the evolution of the inner component. Next, the size of the largest objects was also set arbitrarily. However, decreasing it would make it difficult to explain why destructive collisions are happening at all. Indeed, at least $\sim 100\km$-sized bodies have to have formed in the discs to provide a sufficient level of stirring to trigger the dust production \citep{kenyon-bromley-2008}. 
Increasing the largest size would not help either, because~--- as discussed above~--- the objects larger than $\sim 100\km$ have collisional lifetimes longer that $\sim \Gyr$ and so do not contribute significantly to the dust production.
Finally, many simplifications about the mechanical properties and the collisional outcomes were made, e.g. only destructive collisions were considered.
However, we were able to reproduce the vast 
majority of the systems with these assumptions, which suggests that the model used is a reasonable approximation to the reality.
In summary, a departure from the assumptions we made is either hard to justify or does not help to change the results. So 
there is a reason to believe a few problematic cases are real.

We conclude that the systems in question may indeed require a mechanism other than a local dust production in an asteroid belt analogue to be explained.
Which mechanism may be at work? Comparing the positions of the five systems in Fig.~\ref{fig:KWy14plot2}
with the \citet{kennedy-piette-2015} lines in Fig.~\ref{fig:ParaPlots} 
readily shows that the transport scenario is not a viable explanation.
A number of other possibilities have been proposed \citep[see Section 7 in][for a list of conceivable scenarios]{matthews-et-al-2013}.
These include, for instance, recent major collisions of planetesimals producing dust 
\citep{song-et-al-2005}.
Another possibility is a sudden large influx of comets scattered inwards as a result of dynamical instability of giant planets, similar to the Late Heavy Bombardment in the Solar System \citep{wyatt-et-al-2007,booth-et-al-2009}.
However, warm dust in large amounts observed in such systems does not necessarily have to be a transient phenomenon.
Instead, it may be a sign of a nearly constant influx of comets from the outer belts occurring over long periods of time in mature planetary systems.
Different ways of how multiple planets can scatter comets into the inner systems 
have been discussed, \revised{e.g., in}
\citet{bonsor-wyatt-2012}, \citet{bonsor-et-al-2012}, \citet{faramaz-et-al-2017}.
Once in the inner system, these comets
produce dust through \revised{disintegration, sublimation,} and, if they are present at sufficiently high density,
also collisions. A major \revised{difference from the steady-state two-belt scenario tested in our paper
is that the outer belt in this mechanism} continuously replenishes the inner cometary one.
Another possibility would be an extended disc of comets in very eccentric orbits with periastra in the region
where warm dust is observed \citep{wyatt-et-al-2010}. Such a disc (``mini-Oort cloud'') could, in principle, arise
from planet-planet scattering \citep{raymond-armitage-2013}.

The ``abnormal'' systems are older than $1\Gyr$, their inner components lie at $\sim 2$--$3\AU$, 
and have a fractional luminosity of 
\revised{$(3$--$8)\times10^{-4}$}.
The fraction of these systems in our sample is 5 out of 225, or roughly $2\%$. This is  in agreement 
with other studies suggesting atypically bright warm dust to be found in about $\sim 2\%$ of systems
\citep{bryden-et-al-2006,wyatt-et-al-2007,ishihara-et-al-2016}. Four of the five systems in question are reported here as being exceptional for the first time.
The only one where the warm component was previously identified as atypically bright is HIP~89770 (HD~169666, \citeauthor{moor-et-al-2009} \citeyear{moor-et-al-2009}).
In our sample there are no other discs that are considered as outliers by other studies except for   $\zeta$ Lep \citep{moerchen-et-al-2007} and $q^1$~Eri. Both of them are not classified as exceptions here, since they were fitted well, which can be attributed to the difference in the models used. In the $q^1$~Eri case where previous studies suggested transient or cometary scenarios, \citet{schueppler-et-al-2016} showed that applying a more accurate model reinforces the standard evolutionary scenario. Similar arguments may apply to $\zeta$ Lep. Both systems are an example as to how a different model can change the interpretation of the observations and how important detailed modelling is.

\section{Conclusions}

Many, perhaps the majority, of the debris discs detected so far show signs of a two-component structure, 
comprising colder dust from massive Kuiper-belt analogues and warm dust at a few AU from the stars. The origin of the latter is unknown. In this paper, we attempted to determine whether the warm dust can be locally sustained by asteroid-belt analogues in the process of their steady-state collisional grinding. We assumed that both an ``asteroid belt'' supplying warm dust and a ``Kuiper belt'' responsible for the cold dust population are descendants of an initially extended protoplanetary disc, which was separated into two distinct belts by one or another physical process (for instance, gravity of giant planets). Both belts were assumed to collisionally evolve in a steady-state regime.

To implement the scenario described above we parameterized the progenitor protoplanetary disc with two parameters, the total mass (or the density at 1~AU) and the slope of the surface density profile. That disc was evolved with the aid of the analytic model of \citet{loehne-et-al-2007}, predicting the amount of dust expected to remain in the system (at its age) at any distance from the star. This model was then applied to a set of 225 systems from the Spitzer/IRS catalogue \citep{chen-et-al-2014} that likely possess a two-component structure. For each of those systems, we tried to find the parameters that a protoplanetary disc would have to have in order to produce the warm and cold components with the fractional luminosities 
and radii inferred from the observations.

Our findings are as follows:
\begin{enumerate}
\item
We found that the overwhelming majority of the systems (220 of 225, or 98\%) are compatible with the
two-belt scenario and determined the parameters that protoplanetary discs had to have to produce the currently observed two-component debris disc architecture. The average disc mass (taking a standard 100:1 gas-to-dust ratio) was found
to be
$M_\text{PPD}/\revised{\Msun}=\left(3.3^{+0.4}_{-0.3}\right)\times 10^{-3}$, or
$M_\text{PPD}/M_{\star} = \left(2.0^{+0.3}_{-0.2}\right)\times 10^{-3}$,
and a slope of the surface density distribution was derived to be $p=-0.93\pm0.06$. These results are in rough agreement with the masses of protoplanetary discs estimated from their (sub)millimetre observations, as well as with the masses and density profiles of ``minimum-mass extrasolar nebulae'' reconstructions for extrasolar systems with multiple planets.
\item Compatibility with the two-belt, i.e., asteroid- and Kuiper-belt, scenario does not exclude other mechanisms, for instance short-period comets as a source of warm dust. A fraction of these systems may also contain warm dust transported inward from the outer belt by drag forces. A combination of these mechanisms is possible as well.
\item
The remaining five systems are those that clearly harbour two distinct dust components, but cannot
be reproduced with the two-belt scenario. 
These atypical systems are old (\ga $1\Gyr$) and contain warm dust at $\sim 2$--$3\AU$
with a fractional luminosity of  \revised{$(3$--$8)\times10^{-4}$}, which is
too bright to be compatible with a steady-state collisional cascade in an asteroid-belt analogue.
\item
Warm dust in the atypical systems cannot be explained with dust transport from the outer
``Kuiper belts'' either.  Instead, cometary or transient mechanisms must be at work.
Warm dust in these systems could be replenished, for instance, by comets scattered inward from the outer belts or by long-period comets from ``mini-Oort clouds.''
Alternatively, the large amounts of warm dust could be a transient phenomenon,
for instance debris from a recent major collision or a consequence of recent planetary system instability. The fraction of the ``abnormal'' systems in our analysis is $2\%$, close to what was found \revised{in} other studies.

\end{enumerate}

\revised{We are aware that our analysis provides an exceptionally simplified view of the debris discs, both because it utilizes a simple analytic model of disc evolution, and because the sample only includes unresolved discs with a few photometry points from mid- to far-infrared. In reality, the radial structure of discs must be far more complex than just a combination of narrow  ``asteroid'' and ``Kuiper'' belts. For instance, resolved near-inrared interefremetric observations of recent years have shown about 20\% of stars at all ages and spectral classes to possess hot ``exozodiacal'' dust within 1~AU \citep[see, e.g.][and references therein]{absil-et-al-2013,ertel-et-al-2014b}. It remains unclear whether this hot dust is directly related to warm and/or cold belts investigated here. Nor is it clear whether some of these systems are related to transient phenomena. Future studies, involving a detailed analysis of multi-wavelength, resolved data for individual systems, should provide deeper insights into the nature of multi-temperature debris disks.}

\section*{Acknowledgments}
We thank Christine Chen and Mark Booth for a thorough reading of the manuscript prior to submission and a number of valuable comments and suggestions.
\revised{A helpful and speedy review report by Amy Bonsor is greatly appreciated.}
This research was supported by the DFG, grant Kr~2164/15-1.





\appendix
\section{Collisional model}

Here, we summarize the \citet{loehne-et-al-2007} analytic model of the collisional evolution of a debris disc. As in any collisional model, the crucial quantity that controls the collisions is the critical fragmentation energy $Q_\text{D}^{\ast}$, i.e., the minimum impact energy needed to disrupt the target.
It depends on the material and the size of the object and is commonly approximated as the sum of two power laws \citep{benz-asphaug-1999}
\be
Q_\text{D}^{\ast}=A_{\text{s}}\left(\frac{s}{1\m}\right)^{3b_{\text{s}}}+A_{\text{g}}\left(\frac{s}{1\km}\right)^{3b_{\text{g}}} , \label{eq:QD}
\ee
where we chose $b_{\text{s}}=-0.12$, $b_{\text{g}}=0.45$, and $A_{\text{g}}=A_{\text{s}}=5\times10^6 \erg\g^{-1}$, close to the reference values for basalt \citep{benz-asphaug-1999}.

The first term describes the material strength of the object that decreases with the size. The second describes the strength of the object gained through self-gravity and increases with size.
This classifies all solids into two populations, the small grains kept together by the material strength and the large objects consolidated by self-gravity.
One also introduces the
breaking size $s_{\text{b}}$, the size dividing the two populations. It is the size at which both effects, material strength and self gravity, contribute equally to $Q_\text{D}^{\ast}$.

The strength and gravity regimes would set two equilibrium slopes in the size distribution of solids.
The particles in the strength regime, $s_\text{min} < s < s_\text{b}$, have the distribution 
$n(s)\propto s^{3q_\text{s}-2}$.
The objects in the gravity regime, $s_\text{b} < s < s_\text{max}$, follow the distribution
$n(s)\propto s^{3q_\text{g}-2}$.
Here, $s_\text{min}$ is the radius of the smallest grains that can stay in bound orbits against radiaion pressure 
\citep{burns-et-al-1979}, while $s_\text{max}$ is the radius of the largest objects in the disc. 
The slopes ($q$'s) are related to $b$'s as \citep{o'brien-greenberg-2003}
\be
q=\frac{11+3b}{6+3b} .
\ee
The above treatment is still incomplete, as it tacitly assumes that all solids in the disc are involved in the collisional cascade at all times. 
However, different-sized objects in the disc have different collisional lifetimes  
\citep{wyatt-et-al-1999,wyatt-dent-2002}. As a consequence, the largest objects above a certain transitional size $s_t$ 
may have collisional lifetimes that are longer than the current age of the system and thus are not yet involved in the 
cascade. As time elapses, that transitional size $s_t$ increases. 
The objects larger than $s_t$ are assumed to preserve the primordial size distribution slope:
$n(s)\propto s^{3q_\text{p}-2}$. Following \citet{loehne-et-al-2007}, we chose $q_\text{p}=1.87$.

These considerations can be quantified as follows.
We assume that all grains of a specific size enter the collisional cascade at the same time. This point in time is given by \citep{wyatt-et-al-2007}
\be
t(s)=\frac{4\pi}{\sigma_{\text{tot}}} \left(\frac{s} {s_{\text{min}}}\right)^{3q_\text{p}-5}\frac{r^{5/2}\total r}{(\mathcal{G} M_{\star})^{1/2}}\frac{I} {f(e,I)G(q_\text{p},s)} ,\label{eq:time}
\ee
where
$\sigma_{\text{tot}}$ is the total initial cross section of the disc,
$r$ the distance to the star,
$\total r$ the disc width,
$e$ and $I$ the orbital eccentricity and inclination of objects (we assume $e=I/2$),
$s_{\text{min}}$ the smallest grain size,
$\mathcal{G}$ is the gravitational constant,
and $M_{\star}$ the stellar mass.
The function $f$ is given by $f(e,I)=(1.25e^2+I^2)^{1/2}$ and $G(q,s)$ is taken from \citet[][Eq. (24)]{loehne-et-al-2007}.
The latter contains an important factor $X_\text{c}(s)$, which is the ratio between the size of the smallest projectile that is able 
to disrupt the target, and the size of that target $s$: 
\be
X_\text{c}(s)=\left(\frac{2 r Q_\text{D}^{\ast}(s)}{f(e,I)^2~\mathcal{G} M_{\star}}\right)^{1/3}.\label{eq:Xc}
\ee

Using Eq.~(\ref{eq:time}), we can determine the maximum size of the objects that have entered the collisional cascade in a system with the age $t$, i.e., compute the size $s_\text{t}$ that separates the primordial and evolved distribution.

Combining the above ingredients, the entire size distribution in the model of \citet{loehne-et-al-2007} takes the form:
\bea
n(s,t) &=& n_\text{max} \left(\frac{s_\text{max}}{s}\right)^{3q_\text{p}-2}
\label{eq:sizdisp}
\\
&& \text{for} \quad s_\text{t} < s < s_\text{max},
\nonumber\\
n(s,t) &=& n_\text{max} \left(\frac{s_\text{max}}{s_\text{t}}\right)^{3q_\text{p}-2}
\left(\frac{s_\text{t}}{s}\right)^{3q_\text{g}-2}
\label{eq:sizdisg}
\\
&& \text{for} \quad s_\text{b} < s < s_\text{t},
\nonumber\\
n(s,t) &=& n_\text{max} \left(\frac{s_\text{max}}{s_\text{t}}\right)^{3q_\text{p}-2}
\left(\frac{s_\text{t}}{s_\text{b}}\right)^{3q_\text{g}-2}
\left(\frac{s_\text{b}}{s_\text{min}}\right)^{3q_\text{s}-2}
\label{eq:sizdiss}
\\
&& \text{for} \quad s_\text{min} < s < s_\text{b} ,
\nonumber
\eea
assuming that $s_\text{t}>s_\text{b}$.
If $s_\text{t}\le s_\text{b}$, then
only the objects in the strength regime are involved in the cascade, and 
$s_\text{b}$ is replaced by $s_\text{t}$.
If $s_\text{t} \ge s_\text{max}$, then
the largest objects have already entered the collisional cascade,
and $s_\text{t}$ is replaced by $s_\text{max}$.
At this late stage in the evolution
the depletion of the disc becomes apparent.
Since no new material enters the cascade, while the smallest grains are expelled from the system by radiation pressure, 
the fractional luminosity of the disc declines.
This is described through the time dependence of $n_\text{max}$ in Eqs.~(\ref{eq:sizdisp})--(\ref{eq:sizdiss}):
\be
n_\text{max} = \frac{n_\text{max}(0)}{1+t/t_\text{max}}\label{eq:nmax}
\ee
where $n_\text{max}(0)$ is the amount of objects of the largest size at the beginning of the evolution and
the collisonal lifetime of the largest objects, $t_\text{max}$, can be calculated from Eq.~(\ref{eq:time}) by inserting 
$q_\text{g}$ for $q_\text{p}$ and $s_\text{max}$ for $s$.

Using the size distribution Eqs.~(\ref{eq:sizdisp})--(\ref{eq:sizdiss}),
the dust luminosity at any given time in the evolution can expressed as
\be
f_\text{d}(t)=\frac{\sigma(t)}{4\pi ~r^2}
\ee
and
\be
\sigma(t)=\pi~\int_{s_\text{min}}^{s_\text{D}}n(s,t)~s^2\total s ,
\ee
where only the grains smaller than $s_\text{D}$ (set to $1\mm$) are assumed to contribute to the cross section.
Expressing $n_\text{max}(0)$ through the initial mass $M_{\text{0}}$ of the evolving disc
\citep[][Eq.~(29)]{loehne-et-al-2007} yields the formula for the fractional luminosity in the final form:
\be
\begin{split}
f_\text{d}(t) = \frac{6-3q_\text{p}}{16\pi\rho r^2s_\text{max}^4(q_{\text{s}}-5/3)}
\left[1-\left(\frac{s_{\text{max}}}{s_{\text{min}}}\right)^{3q_\text{p}-6}\right]^{-1}\\ 
\times\frac{M_0 s_{\text{b}}^3}{1+t/t_{\text{max}}}
\left(\frac{s_{\text{max}}}{s_{\text{t}}}\right)^{3q_\text{p}-2}
\left(\frac{s_{\text{t}}}{s_{\text{b}}}\right)^{3q_{\text{g}}-2}\\
\times\left[\left(\frac{s_{\text{b}}}{s_{\text{min}}}\right)^{3q_{\text{s}}-5}-\left(\frac{s_{\text{b}}}{s_\text{D}}\right)^{3q_{\text{s}}-5}\right] ,
\end{split}\label{eq:fdt}
\ee
with $\rho$ being the bulk density of the material.
Similar to Eqs.~(\ref{eq:sizdisp})--(\ref{eq:sizdiss}),
this equation is valid for
$s_\text{b}< s_\text{t} < s_\text{max}$.
Both at early and late stages of the evolution, some modifications are required.
These are listed in Tab.~\ref{tab:changes}.
\begin{table}
\caption{
Changes to Eq.~\ref{eq:fdt} depending on the evolutionary stage.\label{tab:changes}}
\begin{center}
\begin{tabular}{lll}
\hline
Stage of evolution & Replace & With \\
$s_\text{t}\le s_\text{b}$ & $s_\text{b}$ & $s_\text{t}$\\
$s_\text{b}< s_\text{t} < s_\text{max}$ & - & -\\
$s_\text{max} \le s_\text{t}$ & $s_\text{t}$ & $s_\text{max}$\\
\hline
\end{tabular}
\end{center}
\end{table}
\label{lastpage}

\bsp
\end{document}